\documentclass{PoS}
\usepackage{amsmath,amssymb}

\newcommand{\tmtextbf}[1]{{\bfseries{#1}}}

\newcommand{\tmtexttt}[1]{{\ttfamily{#1}}}

\newcommand{\bea}{\begin{eqnarray}}
\newcommand{\eea}{\end{eqnarray}}
\newcommand{\eq}[1]{Eq.~(\ref{#1})}
\newcommand{\fig}[1]{Fig.~\ref{#1}}
\newcommand{\ev}[1]{\langle #1 \rangle}
\newcommand{\re}{{\rm e}}

\newcommand{\rD}{{\rm D}}
\newcommand{\Seff}{S_{\rm eff}}
\newcommand{\oV}{\overline V}
\newcommand{\oH}{\overline H}

\newcommand{\ov}{\overline v}
\newcommand{\Sw}{S_{\rm w}}

\newcommand{\Real}{{\rm Re}\,}
\def\tr{{\rm tr}}

\title{Dimensional reduction and confinement from five dimensions}

\ShortTitle{Dimensional reduction and confinement from five dimensions}

\author{\speaker{Francesco Knechtli}, Nikos Irges\footnote{
        Present address: Department of Physics, National Technical University
        of Athens, GR-157 80 Zografou, Attikis Greece} and Antonio Rago\\
        Department of Physics, Bergische Universit\"at Wuppertal\\
        Gaussstr. 20 \\ 
        D-42119 Wuppertal, Germany \\
\\
        E-mail: \email{knechtli@physik.uni-wuppertal.de}}

\abstract{
We study non-perturbatively five-dimensional $SU(2)$ gauge theories by means of
the mean-field expansion on the lattice. On the anisotropic torus we show
that a continuum limit can be defined where the anisotropy is a relevant
parameter. The analysis of the static force supports the fact that the
four-dimensional hyperplanes decouple from each other in the continuum limit.
Clear signs of confinement are found in the static potential along the
hyperplanes.
We present first results from Monte Carlo simulations on the phase diagram.
\begin{flushright} WUB/10-27 \end{flushright}
}

\FullConference{The XXVIII International Symposium on Lattice Field Theory, Lattice2010\\
		June 14-19, 2010\\
		Villasimius, Italy}
\begin{document}
\section{Introduction and conclusions}

Five-dimensional gauge theories are known to be perturbatively trivial, i.e.
removing the ultraviolet cut-off leads to a free theory of photons
\cite{Dienes:1998vg}.
The epsilon expansions \cite{Gies:2003ic,Morris:2004mg}
suggests that there might be a 
non-perturbative ultraviolet fixed point, where the cut-off could be removed
yielding an interacting theory, but this has been elusive so far in Monte
Carlo simulations. Recently \cite{Irges:2009bi,Irges:2009qp}
a ultraviolet fixed point was found within a mean-field computation.
The computation is
performed using the regularization of a pure $SU(2)$ gauge theory
on an anisotropic Euclidean lattice and
the continuum limit can be taken numerically at fixed physical box size and
fixed anisotropy $\xi$. The latter is defined as the ratio of the
lattice spacing along the usual four dimensions and the one along the
extra dimension, $\xi=a_4/a_5$. The continuum limit can be taken when $\xi<1$,
a region of phase space that was not yet explored.

Monte Carlo simulations of five-dimensional $SU(2)$ gauge theories
concentrated so far in the region where $\xi\gg1$
\cite{Ejiri:2000fc,deForcrand:2010be},
where dimensional reduction from five to four
dimensions is expected to occur because the fifth dimension compactifies.
The authors of \cite{Ejiri:2000fc} provided
numerical evidence for the existence of a critical
radius $R_c\approx0.3/\sqrt{\sigma}$ of the extra dimension
($\sigma$ is the string tension), below which the theory is four-dimensional.

In this talk we will concentrate on the regime $\xi<1$.
We first review the results of the mean-field computation
adding to the data of \cite{Irges:2009qp} results from a larger lattice.
Dimensional reduction seems to happen due to a localization
mechanism. We will then present the exploration of the phase diagram through
Monte Carlo simulations. So far we located a line of first order bulk phase
transitions. It has been recently claimed in \cite{Farakos:2010ie} that this
phase transition could be of second order for parameters that are outside the
range we investigate in this talk (see \fig{f_pd}).
We also see second order phase transitions
due to compactification of one of the usual four dimensions.
Detailed results of our Monte Carlo simulations will be presented
in a forthcoming publication \cite{FMA}. 

\section{The mean-field laboratory for five-dimensional gauge theories
\label{s_mf}}

The mean-field expansion for gauge theories is reviewed in 
\cite{Drouffe:1983fv}.
Path integral expectation values of operators ${\cal O}$ over
$SU(N)$ gauge links $U$ with action $S_G[U]$ are replaced by integrals
over $N\times N$ complex matrices $V$ and Lagrange multipliers $H$
\bea
\ev{{\cal O}[U]} & = &
\frac{1}{Z} \int \rD V \int \rD H \, {\cal O}[V] \re^{-\Seff[V,H]} \\
\Seff & = & S_G[V] + u(H) + (1/N)\Real\tr\{HV\} \,, \\
\re^{-u(H)} & = & \int \rD U \, \re^{(1/N)\Real\tr\{UH\}} \,.
\eea
The mean-field saddle point (or background) is defined by the minimization of
the classical effective action in terms of constant fields 
\bea
H\longrightarrow \oH\mathbf{1} \,;&
V\longrightarrow \oV\mathbf{1} \,;&
\Seff[\oV,\oH]\;\mbox{=minimal} \,. \label{mfbg}
\eea
Gauge invariance is not broken, since arbitrary gauge transformations of the 
constant background solution \eq{mfbg} also fulfill the saddle point equations
\cite{Drouffe:1983fv}. 
Corrections are calculated from Gaussian fluctuations around the saddle point
solution
\bea
H = \oH + h \;& \mbox{and}\; & V = \oV + v \,.
\eea
We impose a covariant gauge fixing on $v$. In \cite{Ruhl:1982er} it was shown
that this is equivalent to gauge-fix the original links $U$.

Our setup is a
$SU(2)$ gauge theory formulated on a 
$L_T\times L^3\times L_5$ Euclidean lattice
with anisotropic Wilson plaquette action \cite{Wilson:1974sk}
\bea
\Sw[V] =
\frac{\beta}{4} \left[ \frac{1}{\gamma}\,\sum_{\rm 4d-p} 
\tr\left(1-V(p)\right) + \gamma\,\sum_{\rm 5d-p} 
\tr\left(1-V(p)\right) \right] \,,
\eea
where the sums run over all the four-dimensional and separately all the
five-dimensional plaquettes $p$ counted with both orientations and $V(p)$
is the product of the link variables $V$ around the plaquette $p$.
The dimensionless bare gauge coupling $g_0$ is defined through $\beta=4/g_0^2$.
We include the anisotropy factor $\gamma$.
At tree level in the gauge coupling $\gamma=\xi=a_4/a_5$.
Due to the anisotropy, the mean-field
background is $\ov_0$ along directions $\mu=0,1,2,3$ and $\ov_{05}$ along the
fifth dimension. We compute the following observables to leading order in the
mean-field expansion \cite{Irges:2009bi}: the static potential $V_4(r)$ along the
four-dimensional hyperplanes orthogonal to the fifth dimension,
the static potential $V_5(r)$ along the fifth
dimension, the scalar (or ``Higgs'') mass $m_H$ and the vector gauge boson
mass $m_W$.
\begin{figure}\centering
  \resizebox{7cm}{!}{\includegraphics{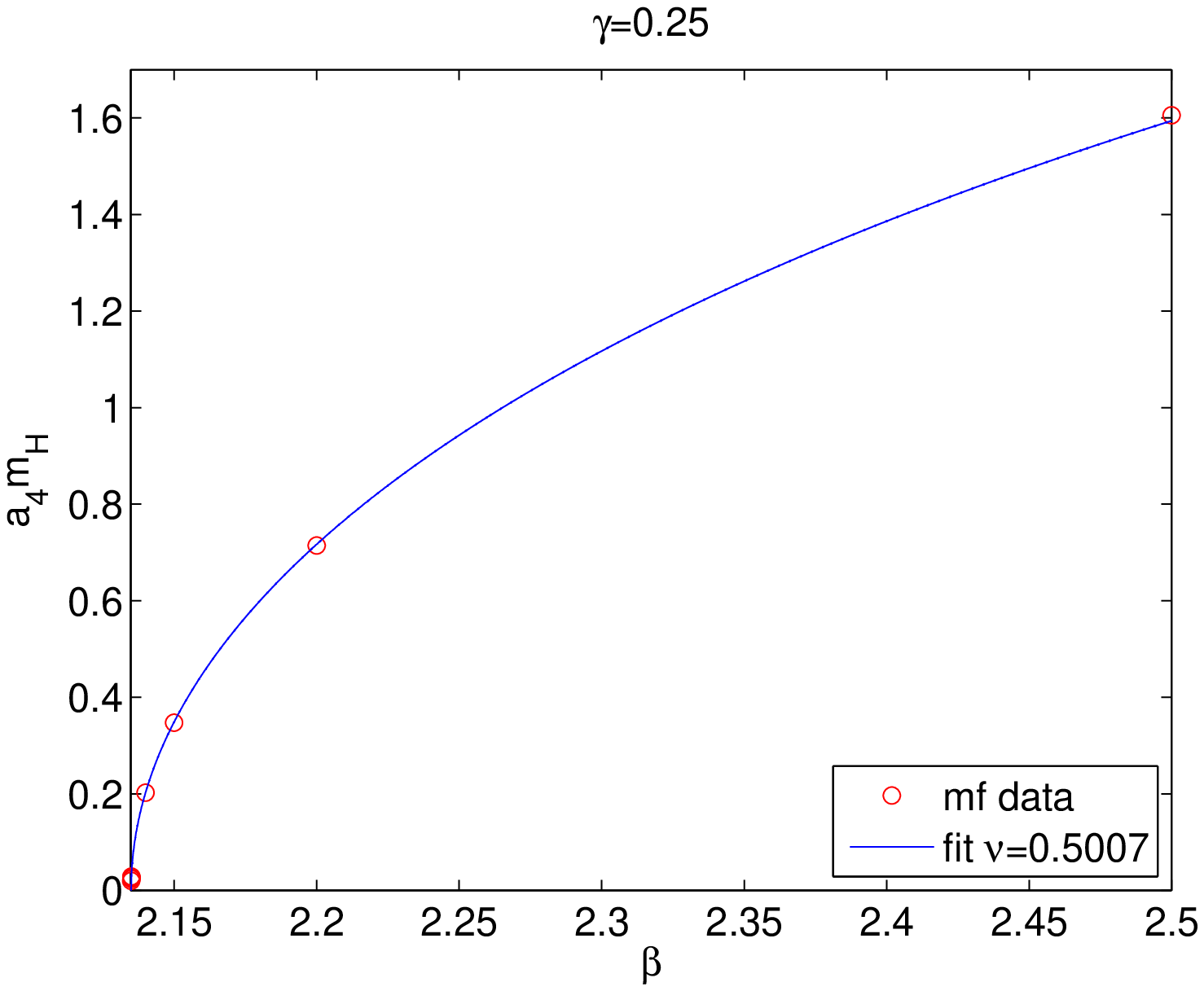}} \ \
  \resizebox{7cm}{!}{\includegraphics{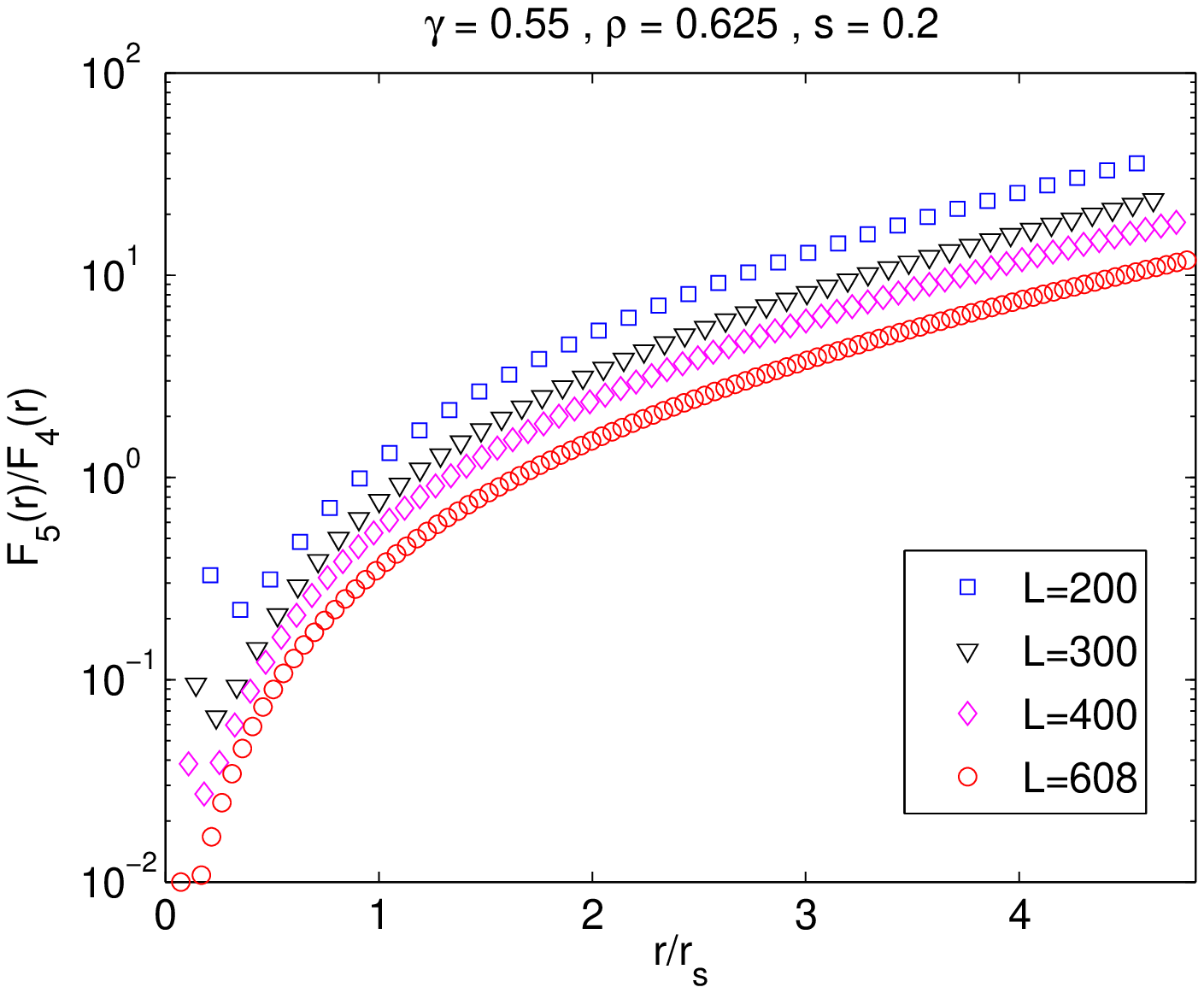}}
  \caption{Left plot: the scalar mass $m_H$
    in the d-compact phase at $\gamma=0.25$. 
    It approaches zero consistently with a 
    second order phase transition at $\beta_c=2.1349$.
    Right plot: the ratio of the force $F_5$ along the fifth dimension to the
    force $F_4$ along the four-dimensional hyperplanes, computed along a LCP.}
\label{f_mH_Fratio}
\end{figure}

\section{Dimensional reduction and continuum limit
\label{s_cl}}

The phase diagram of the theory can be mapped through
the values of the mean-field 
background. There is a confined phase ($\ov_0=0$, $\ov_{05}=0$),
a layered phase ($\ov_0\neq0$, $\ov_{05}=0$)
and a deconfined phase ($\ov_0\neq0$, $\ov_{05}\neq0$).
By looking at the short distance behavior of $V_4$ we can decide, whether
dimensional reduction occurs, in which case the potential has a
four-dimensional Coulomb form $V_4\sim1/r$ (as opposed to a five-dimensional
Coulomb law $V_4\sim1/r^2$). We identify two dimensionally reduced regions in
the deconfined phase \cite{Irges:2009bi}: 
one at $\gamma\gg1$ (compact phase) and one at
$\gamma<1$ close to the phase boundary with the layered phase (we call it the
``d-compact'' phase, where ``d'' stands for dual to the compact phase, since
they are related by $\gamma\to1/\gamma$). 
In the following, we study the properties of the d-compact phase.
\begin{figure}[p!]\centering
  \resizebox{7cm}{!}{\includegraphics{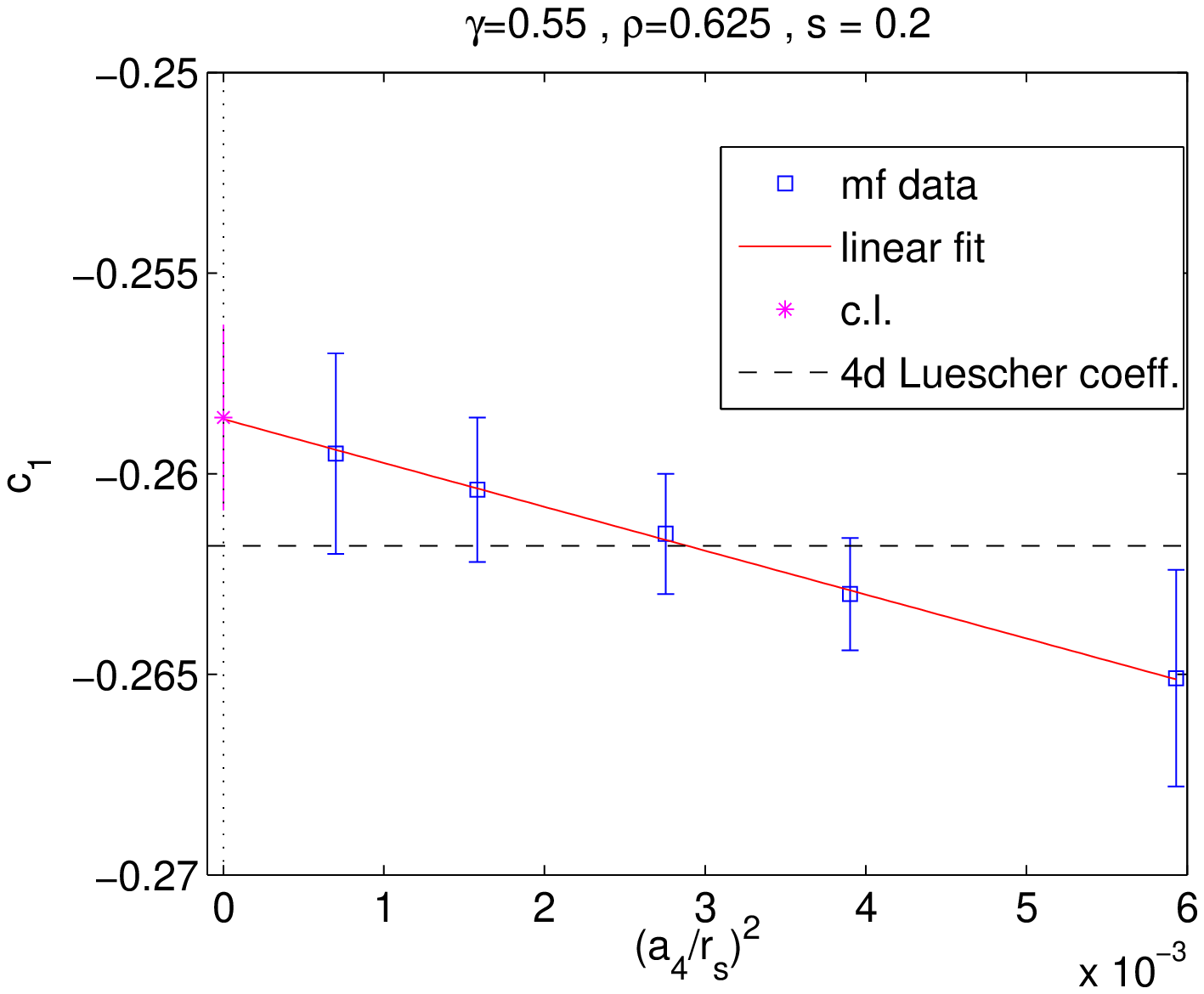}} \ \
  \resizebox{7cm}{!}{\includegraphics{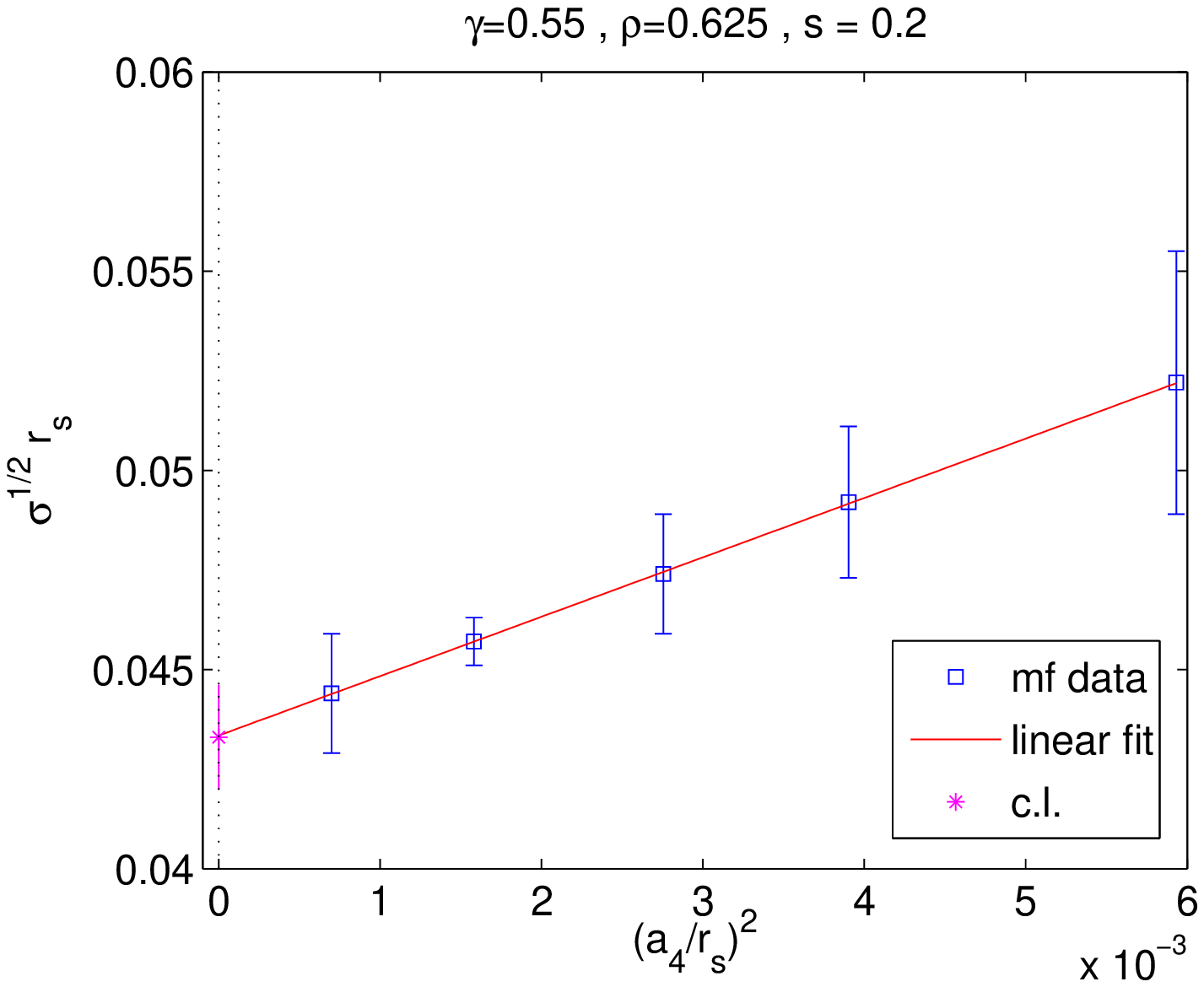}}
  \caption{
    Continuum extrapolations of the coefficient $c_1$ and the string
    tension $\sqrt{\sigma}r_s$. The coefficients are fitted
    locally and are averaged in the common plateau range
    $2.15\le r/r_s\le 2.80$.
} 
\label{f_mf_c1_sigma}
\end{figure}
\begin{figure}[p!]\centering
  \resizebox{7cm}{!}{\includegraphics{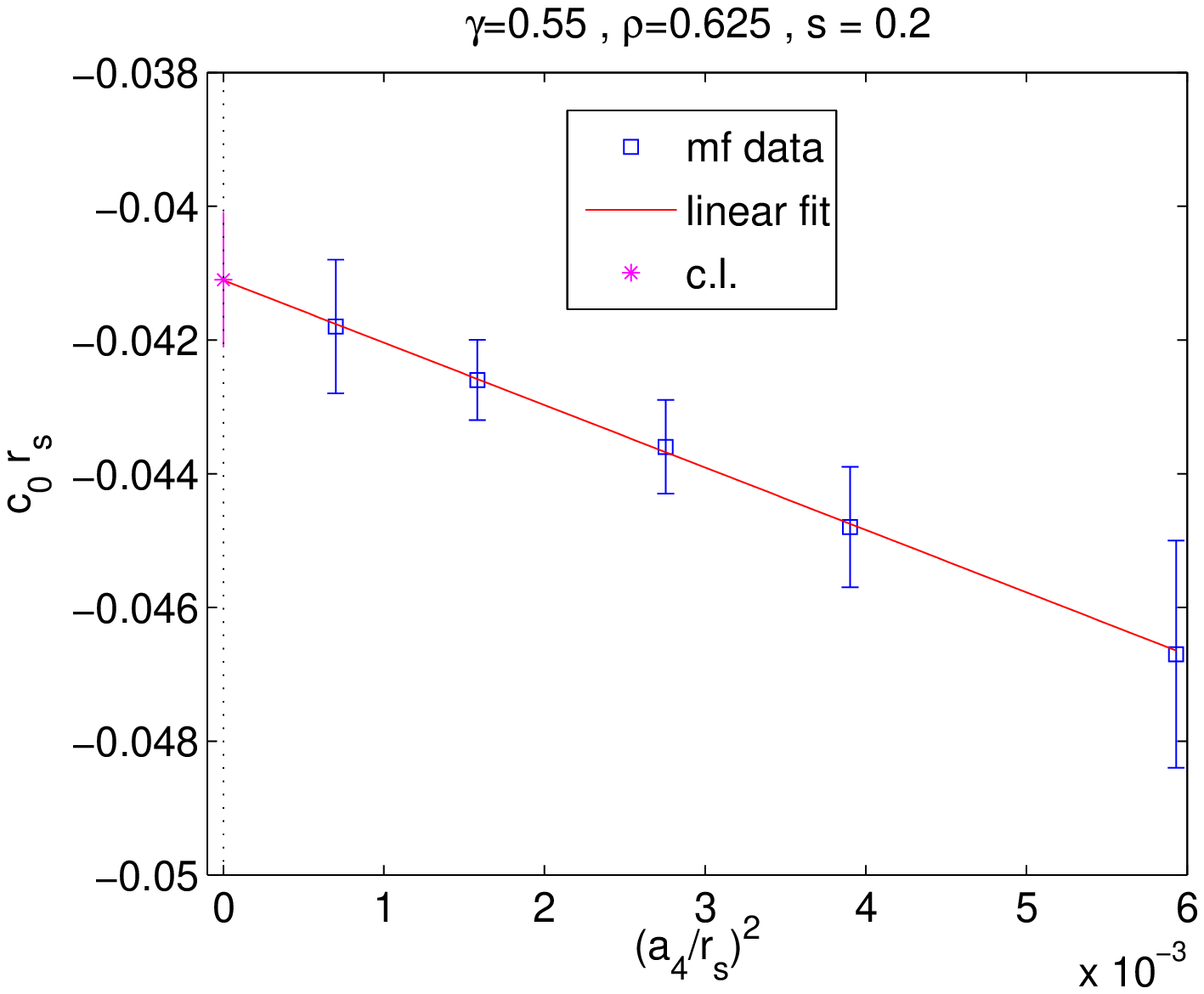}} \ \
  \resizebox{7cm}{!}{\includegraphics{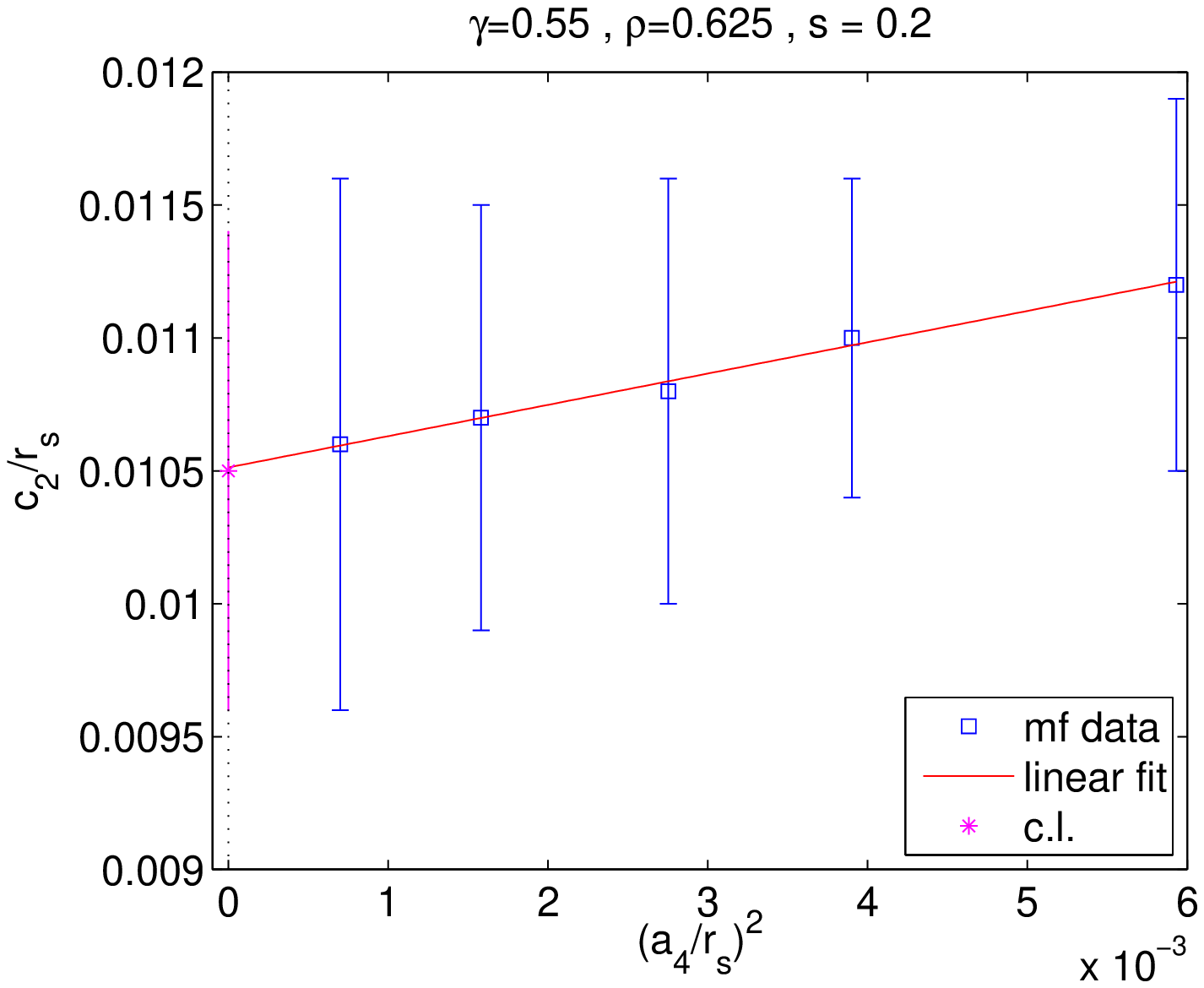}}
  \caption{
    Continuum extrapolations of the coefficients $c_0r_s$ and 
    $c_2/r_s$.
}
\label{f_mf_c0_c2}
\end{figure}

In the left plot of \fig{f_mH_Fratio} we show the results for the scalar
mass $m_H$ at $\gamma=0.25$. It goes to zero when the
phase boundary with the layered phase at $\beta_c=2.1349$ is approached.
It obeys the scaling law
$a_4m_H\propto(1-\beta_c/\beta)^\nu$
of the inverse correlation length of a second order phase transition
with exponent fitted to $\nu=0.5007$. The value $\nu=1/2$ is the one
of the four-dimensional Ising model and is
in agreement with the analysis of \cite{Svetitsky:1982gs}.
The vector gauge boson mass is essentially independent of $\beta$ and $\gamma$
(at leading order in the mean-field expansion) and depends essentially only on
$L$ through $a_4m_W=c_L/L+0.0010(5)$ with $c_L=12.51(1)$. 
This means that in the infinite volume limit the gauge boson is massless,
i.e. no spontaneous gauge symmetry breaking takes place.

The existence of a second order phase transition means that we can take the
continuum limit. We define lines of constant physics (LCP) by keeping
\bea
\gamma=0.55 & \mbox{and} & \rho=m_W/m_H=0.625 \label{LCP}
\eea
constant. We set $L=L_T=L_5$ and take
the continuum limit $L\to\infty$ by computations on a series of lattices
$L=200,300,400,608$.
In the right plot of \fig{f_mH_Fratio} we show the ratio of the static force
$F_5(r)=\{V_5(r)-V_5(r-a_5)\}/a_5$ along the extra dimension to the 
static force $F_4(r)=\{V_4(r)-V_4(r-a_4)\}/a_4$ along the four dimensional
hyperplanes. 
In order to plot physical quantities we need a way
to set the lattice spacing, i.e. a quantity which is in principle measurable
in four dimensions. Since the
vector boson is massless, we cannot take $a_4m_W$. Instead we define a scale
$r_s$ similarly to \cite{Sommer:1993ce} from the static force $F_4$
through the equation $r^2\,F_4(r)|_{r=r_s}=s=0.2$. In \fig{f_mH_Fratio} we
plot the force ratio as a function of $r/r_s$. We observed that
the $r^2\,F_4$ has a finite continuum limit \cite{Irges:2009qp}. The data
shown in \fig{f_mH_Fratio} suggests that $r^2\,F_5$ goes to zero as
$L\to\infty$. The interpretation of this result is
that the system reduces in the continuum to a set of non-interacting
four-dimensional hyperplanes. Dimensional reduction occurs through
localization. 

Since at short distance the static potential $V_4$ has a four-dimensional
Coulomb form (see \cite{Irges:2009qp}), we fit it for $r/r_s>1$ on the
lattices along the LCP \eq{LCP} to the form
\bea
V_4(r) = \mu + \sigma r + c_0 \log(r) + \frac{c_1}{r} + \frac{c_2}{r^2}\,. 
\label{V4fit}
\eea
We do the fits locally, i.e. the coefficients $\sigma$, $c_0$, $c_1$ and $c_2$
are functions of $r$.
In the range $2.15\le r/r_s\le 2.80$ the coefficients have {\em simultaneous}
plateaus, which we plot in \fig{f_mf_c1_sigma} and \fig{f_mf_c0_c2} together
with their continuum extrapolations linear in $(a_4/r_s)^2$. We make the
coefficients dimensionless by appropriate powers of $r_s$. The continuum limit
of the coefficient $c_1$ is consistent with the universal value of the
L\"uscher term $-(d-2)\pi/24$ \cite{Luscher:1980fr,Luscher:1980ac} in $d=4$.
This is our strongest evidence for dimensional reduction. In the continuum, we
get a positive string tension together with a
large and negative logarithmic correction. We cannot explain at
the moment the origin of the negative logarithmic term, it hints to a
dimensionally reduced theory in four dimensions different than a pure gauge
theory, where such a term does not occur. 
\begin{figure}\centering
  \resizebox{12cm}{!}{\includegraphics[angle=-90]{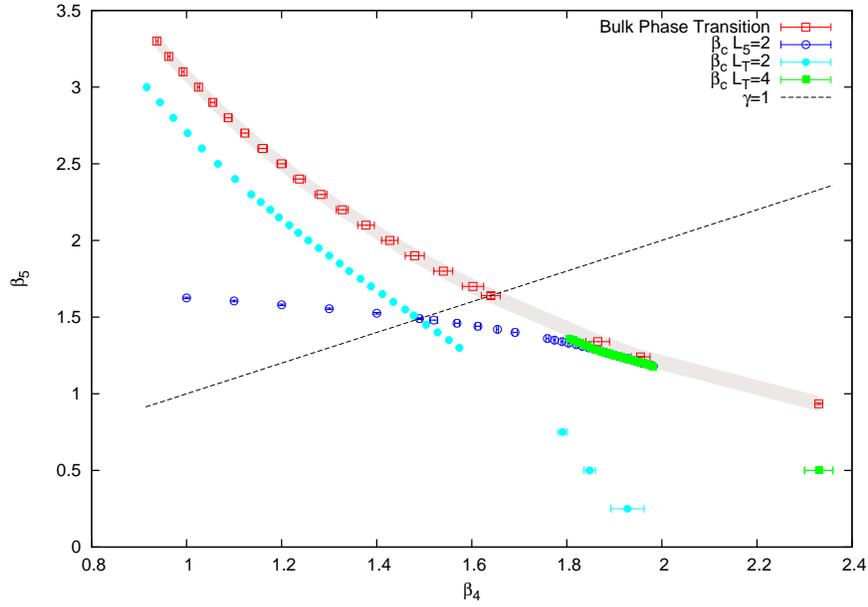}}
  \caption{The phase diagram of the five-dimensional $SU(2)$ gauge theory
    based on Monte Carlo simulations.}
  \label{f_pd}
\end{figure}

\section{The phase diagram from Monte Carlo simulations
\label{s_mc}}

The mean-field calculation gives a consistent picture of a second order phase
transition at anisotropy parameter $\gamma<1$. The next question is if this
picture is validated by Monte Carlo simulations.

In \fig{f_pd} we summarize our Monte Carlo results for the phase diagram. We
denote by symbols the location of phase transitions in the
$\beta_4=\beta/\gamma$ and $\beta_5=\beta\gamma$ plane.
The red empty squares and the shaded line denote the bulk phase transition,
which was first studied in \cite{Creutz:1979dw}. 
It separates a confined phase, where the expectation values of the Polyakov
lines are zero, from a deconfined phase, where they are non-zero along every
direction and regardless of the lattice size.
The bulk phase transition is of first order everywhere
in the range of parameters plotted. This is signaled by a hysteresis curve
in the plaquette observables and by a double peak in its susceptibility (when
plotting simulation results from both cold and hot starts). In order to see
this hysteresis the lattice volume has to be large enough and this is an issue
in particular for the parameter region where $\gamma<1$ (or
$\beta_4>\beta_5$), since there one has to make the number of points in four
directions large.
\begin{figure}\centering
  \resizebox{7cm}{!}{\includegraphics{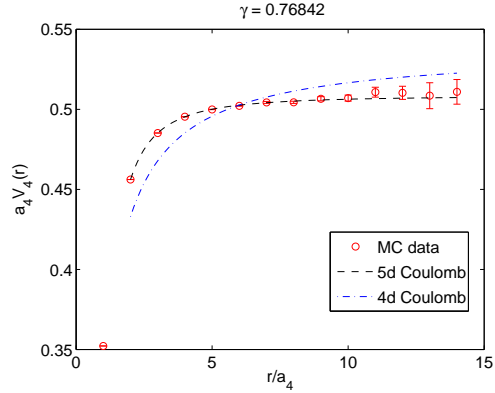}}
  \caption{The static potential $V_4$ on a $32^4\times L_5=16$ lattice in the
    deconfined phase at $\beta_5=1.24$, $\beta_4=2.10$.}
  \label{f_pot}
\end{figure}

We study dimensional reduction by computing the static potential from
Wilson loops. We distinguish two classes of Wilson loops, the ones in the
four-dimensional hyperplanes (in directions $(t,x)$, $(t,y)$ and $(t,z)$) and
the ones along the fifth dimension in directions $(t,5)$. From the former we
extract the static potential $V_4$, from the latter $V_5$.
In computing the Wilson loops, we use the
multi-hit (or one-link) method \cite{Parisi:1983hm} for the temporal links and
two levels of HYP smearing \cite{Hasenfratz:2001hp} in directions $(x,y,z,5)$
for the spatial links. As an example, we show in \fig{f_pot} preliminary data
for the static potential $V_4$ in the deconfined phase at $\gamma<1$.
It can be perfectly fitted by a five-dimensional Coulomb potential.

At present, the only second order phase transitions that we could locate
correspond to finite temperature (compactification). At $\gamma>1$ we confirm
the results of \cite{Ejiri:2000fc}. For $\gamma<1$ the situation is shown in
\fig{f_pd}, in the region below the dashed line corresponding to $\gamma=1$ .
We located lines of phase transitions due to compactification of the fifth
dimension or of the temporal dimension, signaled
by a peak in the susceptibility of the respective Polyakov loop.
As they evolve in the parameter space these lines cannot cross the line of
the bulk transition, they will instead accumulate on it becoming always of
first order.
For the case of compactification with $L_5=2$ points (blue empty circles) or
of with $L_T=2$ points (cyan filled circles),
the finite temperature transition is always of second order if it
happens far enough from the bulk phase transition.
The order with $L_T=4$ points (green filled squares) is still under
investigation. We will report in detail in \cite{FMA}.

\acknowledgments

N.Irges thanks the Alexander von Humboldt Foundation and
A. Rago the German Science Foundation (DFG) for support.
The Monte Carlo simulations were carried out on the
Cheops supercomputer at the RRZK computing centre of the University of Cologne
and on the cluster Stromboli at the University of Wuppertal and we thank both
Universities.

\end{document}